\documentclass[prl,twocolumn,showpacs,showkeys]{revtex4}

\usepackage{graphicx}% Include figure files
\usepackage{dcolumn}% Align table columns on decimal point
\usepackage{bm}% bold math
\usepackage{amssymb}

\begin{document}
%\preprint{}
%\tightenlines
\title{The photon magnetic moment has not a perpendicular component and is fully paramagnetic}
\author{ H. P\'erez Rojas and E. Rodriguez Querts}
\affiliation{Instituto de Cibernetica, Matematica y Fisica, Calle E
309, Vedado, Ciudad Habana, Cuba.\\}
\date{\today}

\begin{abstract}
Our paper Phys. Rev. D \textbf{79}, 093002 (2009), in which it was
shown the paramagnetic behavior of photons propagating in magnetized
vacuum, is criticized  in Phys. Rev. D \textbf{81}, 105019, (2010)
and even claimed that the photon has a diamagnetic component. Here
it is shown  that such criticism is inadequate and  that the alleged
``perpendicular component" is due to a mistake in differentiating a
vanishing term with regard to the magnetic field $B$, or either by
mistaking the derivative of a scalar product as that of a dyadic
product. A discussion on the physical side of the problem is also
made.
\end{abstract}



\maketitle

 \section{Introduction}
We have shown in \cite{EliPRD} that for a photon moving in a
magnetic field $\mathbf{B}$, assumed constant and homogeneous,(and
for definiteness, taken along the $x_3$ axis, thus $|B_3|=B$,
$B_1=B_2=0$), an anomalous magnetic moment defined as
$\mu_{\gamma}=-\partial \omega/\partial B$ arises. This quantity
\textit{has meaning, and can be defined only} when the photon mass
shell includes the radiative corrections, i.e., the magnetized
photon self-energy, and calculated only after the solution of the
photon dispersion equations \cite{shabad2}. It was shown that it is
paramagnetic ($\mu_{\gamma}>0$), since it arises physically when the
photon propagates due to the magnetic response of the virtual
electron-positron pairs of vacuum, under the action of $\mathbf{B}$,
leading to vacuum magnetization. Thus, the photon embodies both
properties of the free photon and of a magnetic dipole, which leads
to consider it more as a quasi-photon, in analogy with the polariton
of condensed matter physics. Such properties are valid in the whole
region of transparency, which is the region of momentum space where
the photon self-energy, and in consequence, its frequency $\omega$,
is real. Such region is defined for photon transverse momentum
$(\omega^2-k_{\parallel}^2)^{1/2}\leq 2m$, where $\omega$,
$k_{\parallel}$, $k_{\perp}$ are the photon frequency and momentum
components along $\mathbf{B}$, and $m$ is the electron rest energy).

As pointed out in \cite{EliPRD}, beyond that region, as the photon
becomes unstable \cite{shabad2} for frequencies $\omega\geq 2m$ (and
has a significant probability of decaying in electron-positron
pairs), the photon magnetic moment loses meaning if considered
independent of the magnetic moment produced by the electron-positron
background.

In a recently published paper, \cite{Selym} some criticism is made
on \cite{EliPRD}. In it is   pointed out that i)the vector character
of $\boldsymbol{\mu}_{\gamma}$ was ignored in \cite{EliPRD}, ii) its
connection with angular momentum was not shown, and iii) that it was
not mentioned an alleged precession of $\boldsymbol{\mu}_{\gamma}$
around $\mathbf{B}$, due to an hypothetical component orthogonal to
$\mathbf{B}$, which would lead to a diamagnetic behavior.

The present comment is devoted  to demonstrate that such criticism
is lacking of any  basis. We will show that on the opposite, some
 results claimed as true in \cite{Selym}, and differing from those of \cite{EliPRD}
\textit{ are a consequence of mistakes done in handling elementary
vector analysis}. These claims also are in full contradiction with
what can be expected from the well known background quantum dynamics
of electrons and positrons in an external constant and uniform
magnetic field \cite{Johnson}. This cannot be bypassed when
interpreting  the consequences of the solutions of the dispersion
equations for the photon in magnetized vacuum obtained in
\cite{shabad2}.

As pointed out earlier, the case  studied in \cite{EliPRD},
\cite{shabad2}, \cite{Selym} is based on the hypothesis of a
constant and homogeneous magnetic field defined by the invariants
${\cal F}=2B^2>0=const$, ${\cal G}=\mathbf{E}\cdot \mathbf{B}=0$.
Expressions for physical quantities as the polarization operator
$\Pi_{\mu \nu}$ depend on scalar quantities such as ${\cal F}=2B^2$,
$k^2$ (the total four-momentum squared) and $k_\mu {\cal F_{\mu
\nu}}^2 k_\nu$. It is not necessary to stress that being scalars,
they do not depend on the direction of the coordinate axis.
Obviously, in a specific problem a specific direction for $B$ must
be chosen. Such direction breaks the spatial symmetry, and for
simplicity, it is taken as coinciding with one of the coordinate
axes.

Being specific with the above mentioned criticism i) ``(the authors)
\textit{ignored that the photon magnetic moment is indeed a
vector}". We used the definition of photon magnetic moment as a
generalization of the definition of this quantity for electrons and
positrons done in \cite{Johnson}. Then $\mu_{\gamma}=-\partial
\omega/\partial B$ is understood as the modulus of a vector along
$\mathbf{B}$ since as $B=\sqrt{\mathbf{B}^2}$, we have $\partial
B/\partial \mathbf{B}=\mathbf{n}_{\parallel}$, where
$\mathbf{n}_{\parallel}$ is a unit vector parallel to $\mathbf{B}$.
In \cite{EliPRD} we did not use the word ``vector" regarding
$\boldsymbol{\mu}_{\gamma}$ as we not use the name ``particle" when
we speak about an electron.

Let us consider the expression for the vector
$\boldsymbol{\mu}_{\gamma}=-\partial \omega/\partial \mathbf{B}$ in
the most general case. For any value of $B$ and independently of the
order considered in the loop expansion for the polarization
operator, the photon anomalous magnetic moment will be shown to be a
vector parallel to $\mathbf{B}$. This can be easily deduced from the
photon dispersion equations. Initially we have seven independent
variables: the four components of $k_\mu$ plus the three components
of $B$ in an arbitrary system of reference. By choosing the field
along a fixed axis, say, $x_3$, its three components are reduced to
one $B={\cal F}/2$. Each of the dispersion equations for the
eigenvalues of the polarization operator $\kappa^{(i)}$ ($i=1,2,3$)
impose an additional constraint, reducing them to four, $B$ plus the
three components of $\textbf{k}$ which are $k_1, k_2$ and $k_3\equiv
k_{\parallel}$. As $\kappa^{(i)}$ depend on the photon momentum
components in terms of the invariant variables
\begin{eqnarray}\label{1}
 z_1&=&(\textbf{k} \cdot
 \textbf{B})^2/\textbf{B}^2-\omega^2=k_\parallel^2-\omega^2,\\
 \nonumber
z_2&=&(\textbf{B} \times {\textbf{k}})^2/\textbf{B}^2\\ \nonumber
&=&\textbf{k}^2-(\textbf{k} \cdot
\textbf{B})^2/\textbf{B}^2=k_\perp^2,
\end{eqnarray}
the dispersion equations can be written as
\begin{equation}
   z_1 + z_2=\kappa^{(i)}(z_1,z_2,B),\hspace{1cm}i=1,2,3. \label{dispeq}
\end{equation}
In terms $z_1, z_2$ the independent variables are reduced to two,
for instance, $z_2$ and $B$, if (\ref{dispeq}) is solved as
$z_1=f(z_2, B)$ \cite{shabad2}. But as $k_{\parallel}$ is a
component of the photon momentum, the dependence of $z_1$ on $z_2$
and $B$ in specific calculations is assumed as being contained on
the photon energy $\omega$. Thus we usually write $\omega^2=
k_{\parallel}^2-f(z_2, B)$. \textit{In other words, in the solution
of each of the dispersion equations one assumes} $\omega^2$
\textit{as a function of the independent variables} $k_1, k_2,
k_{\parallel}$ and $B$.

Thus, it directly follows from (\ref{dispeq}) that
\begin{equation}\label{der1}
\frac{\partial z_1}{\partial B} =\frac{\partial
\kappa^{(i)}}{\partial z_1}\frac{\partial z_1}{\partial
B}+\frac{\partial \kappa^{(i)}}{\partial B},
\end{equation}
 which conduces to
 \begin{equation}\label{der2}
 \frac{\partial z_1}{\partial B}=-2\omega \frac{\partial\omega}{\partial B}
=\frac{\frac{\partial \kappa^{(i)}}{\partial B}}{1-\frac{\partial
\kappa^{(i)}}{\partial z_1}}.
\end{equation}
 Finally we get  the vector photon anomalous magnetic  moment as
\begin{eqnarray}\label{magneticmom}
    \boldsymbol{\mu}_{\gamma}^{(i)}& \equiv & -\frac{\partial \omega}{\partial
    \mathbf{B}}\\ \nonumber &=& -\frac{\partial \omega}{\partial B}\frac{\partial B}{\partial
    \mathbf{B}}\\ \nonumber &=& -\frac{\partial \omega}{\partial B}\mathbf{n}_{\parallel} \\
    &=&\frac{1}{2\omega}\frac{\frac{\partial \kappa^{(i)}}{\partial B}}
    {1-\frac{\partial \kappa^{(i)}}{\partial
    z_1}}\mathbf{n}_{\parallel}.\nonumber
\end{eqnarray}
Thus, it has been proved in the most general case that
$\boldsymbol{\mu}_{\gamma}=-\partial \omega/\partial
    \mathbf{B}=-(\partial \omega/\partial B)\mathbf{n}_{\parallel}$
    is a vector parallel to $\mathbf{B}$.

In \cite{EliPRD} the present authors concentrated their efforts in
calculating the quantity $\partial \omega/\partial B$, since it was
obvious to be the modulus of a vector parallel to $\mathbf{B}$.

 ii) On the second criticism ``\textit{...the photon magnetic moment is indeed a
vector, consequently, the connection between this quantity and its
angular momentum was not presented}". We quote from \cite{EliPRD},
Section II: ``as $\mu_{\gamma}$ depends through $\Pi_{\mu
\nu}(k,k^{\prime}|A^{ext})$ on the sums over infinite pairs of
Landau quantum numbers and spins of the $e^{\pm}$ pairs, it cannot
depend on any specific eigenvalue of angular momentum, spin, or
orbit center coordinates". Thus, the criticism made in \cite{Selym}
is out of place. We want to remark here that the photon magnetic
moment vector must be interpreted as parallel to  the direction
$x_3$, as the eigenvalues of the operators $J_3$, $p_3$ and $S_3$,
which are quantities defined along $\mathbf{B}$.

iii) On the third criticism ``\textit{they did not comment about its
precession} (of $\boldsymbol{\mu}_{\gamma}$)\textit{around the
external field axis}". The  present authors cannot comment about
what they did not found to exist, as is seen from the expression
(\ref{magneticmom}).

In the next section we discuss some  fundamental flaws of
\cite{Selym}.

\section{Transverse
photon magnetic moment. The derivative of zero with regard to B?}
\subsection{Mathematical criticism} We concentrate in this subsection on the mathematical
inconsistence of the claims made in \cite{Selym} about the arising
of a component
 of the photon magnetic moment
$\boldsymbol{\mu}_{\gamma}$ perpendicular to $\mathbf{B}$. We start
from the eq.(36) of that paper, which expressed in terms of the
parameters $\alpha, e, m^2, B$ (where $\alpha$ is the fine structure
constant and $e,m$ the electron charge and mass) read as
\begin{equation}
\omega =|\mathbf{k}|-\frac{\alpha e B k^2_{\perp}}{6\pi m^2
e|\mathbf{k}|}. \label{36}
\end{equation}

The second term ($B$-dependent) from (\ref{36}) is treated
separately in \cite{Selym} and called
\begin{equation}
\texttt{U}=-\frac{\alpha e B k^2_{\perp}}{6\pi m^2 e|\mathbf{k}|},
\label{37}
\end{equation}
\noindent and by a process of symmetrization it is obtained its
eq.(40), ($\mathbf{n}_{\perp}$ is a unit vector perpendicular to
$\mathbf{B}$) which can be written as
\begin{equation}
 \texttt{U}=-\frac{\alpha e B k_{\perp}}{6\pi m^2
e|\mathbf{k}|}[\mathbf{n}_{\parallel}\cdot \mathbf{B}
k_{\perp}-\mathbf{n}_{\perp}\cdot \textbf{B} k_{\parallel}]
\label{40}
\end{equation}
Obviously, eq. (\ref{37}) \emph{must be} equal to eq. (\ref{40}),
since $\mathbf{n}_{\perp}\cdot \mathbf{B}=B\cos\frac{\pi}{2}=0$.
Thus, all the process done from eq. (\ref{37}) to eq. (\ref{40}) is
to add zero to (\ref{37}). \textit{Such vanishing quantity, since it
was missed a factor $\cos \pi/2=0$, is used in \cite{Selym} to
deduce physical consequences from it}. This is absolutely out of
meaning. We want to stress, that from (\ref{40}) and
(\ref{magneticmom}), we have
\begin{equation} \boldsymbol{\mu}_{\gamma}=-\frac{\partial \omega}{\partial
\mathbf{B}}=\frac{\alpha e k_{\perp}^2 \mathbf{n}_{\parallel}}{6\pi
m^2 e|\mathbf{k}|}.\label{41}
\end{equation}
However, in \cite{Selym} it is reported to have obtained two
components for $\boldsymbol{\mu}_{\gamma}$,
\begin{eqnarray}\label{44}
\mu_{\gamma}^{\parallel}&=&\frac{\alpha e B k_{\perp}}{6\pi m^2
e|\mathbf{k}|}\mathbf{n}_{\parallel}k_{\perp}. \\ \nonumber
\mu_{\gamma}^{\perp}&=&-\frac{\alpha e B k_{\perp}}{6\pi m^2
e|\mathbf{k}|}\mathbf{n}_{\perp}k_{\parallel}
\end{eqnarray}
where $\mu_{\gamma}^{\perp}$ is presented as a result of
differentiating the scalar product $\mathbf{n}_{\perp}\cdot
\mathbf{B}= B\cos \pi/2=0$. What is done is equivalent to take the
second term in brackets in (\ref{40}), which is a scalar,  as a
dyadic product $\mathbf{n}_{\perp}\mathbf{B}$ (a second-rank
tensor). From the above arguments the existence of a photon magnetic
moment component $\mu_{\gamma}^{\perp} \neq 0$ is debunked.

The problem of constant magnetic field which we are considering is
valid in  the subset of Lorentz frames moving parallel to the
magnetic field pseudovector $B$, independently of the orientation of
the coordinate axes. We must obtain physically equivalent results in
these frames since we are working in a relativistic quantum field
theory. In such frames, the magnetic field is described by the
spatial (pseudo)vector $\textbf{B}$ and the photon momentum
$\textbf{k}$. Under proper rotations, the scalar product $B_i k_i=
B_i \delta_{ij} k_j= B_i R^t_{ij}R_{jl}k_l$, where $R_{ij}$ is a
rotation matrix, is invariant. Thus
 $B_i k_i= B^{\prime}_i
 k^{\prime}_i$, where $B^{\prime}_i=R_{ji}B_j$ and
 $k^{\prime}_i=R_{ji}k_j$. Similarly $\textbf{B}^2=\textbf{B}^{\prime 2}$ and $\textbf{k}^2=\textbf{k}^{\prime
 2}$.
If $\omega$ is a function of the spatial scalars $k_{\parallel}^2$,
$z_2$ and $B$, it is, as well as its derivative $\partial
\omega/\partial B$, also a scalar. Thus, they
 cannot depend on the orientation of the axes. (Also $B^2$,  $kF^2k$ and $k_{\parallel}^2-\omega^2$ are
 scalars in Minkowski space \cite{shabad2}).

The ``demonstration" made in the Appendix of \cite{Selym} is again
due to a flaw done in misusing the rotational invariance. In
\cite{Selym} it is wanted to show that by rotating the coordinate
axes, and by putting as equal to zero some component of
$\mathbf{B}$, the previous derivative with regard to such component
of $\mathbf{B}$ leads to a vector orthogonal to $\mathbf{B}$.

The procedure followed in \cite{Selym} is equivalent to the
following one: Let us start from the initial expression
$(kF^2k)=|\mathbf{B} \times \mathbf{k}|^2=B^2 k_{\perp}^2$ obtained
when the only component (of the spatial part) of the tensor $F_{\mu
\nu}$, which we name $F_{ij}$, is $F_{12}=-F_{21}=B$. Let us choose
a system of coordinates rotated an angle $\theta$ around the $x$
axis. By doing it we have changed the magnetic field and momentum
components to $B_y=B \sin \theta$,
 $B_z=B\cos \theta$, (we call the tensor with rotated components $F^{\prime}_{ij}$
  and $\mathbf{B}^{\prime}=(0,B_y,B_z)$
 and $k_y^{\prime}=k_y \cos \theta + k_z\sin \theta$
and
 $k_z^{\prime} =-k_y \sin \theta + k_z \cos \theta$). We have

\begin{equation}\label{F}
F_{ij}^{\prime}k_j^{\prime}=((B_zk_y^{\prime}-B_yk_z^{\prime}), -B_z
k_x^{\prime}, B_y k_x^{\prime})
\end{equation}
Rotational invariance demands that $k^{\prime}_j F_{jl}^{\prime 2}
k^{\prime}_l=k_j F^2_{jl}k_l$. Thus,
\begin{equation}\label{kFk}
(kF^2k)=k_x^{\prime 2}(B_z^{2}+B_y^{2})+(B_z k_y^{\prime}-B_y
k_z^{\prime})^2=[{\bf B}^{\prime}\times {\bf
   k}^{\prime}]^2.\label{crit}
\end{equation}
Obviously $B^2=B^{\prime2}$, $k^2=k^{\prime2}$. We easily check that
the second term in (\ref{crit}) is $B^{\prime2} k^{\prime2} \sin^2
\phi$. The contribution of $\mu_{y}=(\partial \omega/\partial
k^{\prime}F^2 k^{\prime})(\partial k^{\prime}F^2
k^{\prime})/\partial B_y$ to the
perpendicular-to-$\mathbf{B}^{\prime}$ magnetic moment would be
proportional to
\begin{equation}\label{18}
   \frac{\partial k^{\prime}F^2k^{\prime}}{\partial
   B_y}=2k_x^{2\prime}B_y-2(B_z k_y^{\prime}-B_y k_z^{\prime})k_z^{\prime}.
\end{equation} It is argued that if it is taken the limit $B_y\rightarrow 0$, it would lead to

\begin{equation}\label{B_y=0}
   \left.\frac{\partial k^{\prime}F^2k^{\prime}}{\partial
   B_y}\right|_{B_y=0}=-2|{\bf B}^{\prime}|k_y^{\prime}k_z^{\prime}.
\end{equation}

It is claimed that a perpendicular component has appeared, but this
is a manifest flaw. The limit $B_y\rightarrow 0$ cannot be taken
arbitrarily since it \textit{violates the rotational invariance of
scalar products}. The problem is changed by doing that. The angle
formed by $\mathbf{B}, \mathbf{k}$, which we will call $\phi$,  is
rotational invariant also and we have, by considering  the first and
last terms of (\ref{crit})
\begin{equation}
(kF^2k)=|\mathbf{B}\times \mathbf{k}|^2=B^2 k^2 \sin^2
\phi=B^{\prime 2} k^{\prime 2}\sin^2 \phi
\end{equation}
As $B^2=\mathbf{B}^{\prime 2}=B_y^2 + B_z^2$, by differentiating
$\mathbf{B}^{\prime 2}$ with regard to $B_y$ what can be obtained is
$\partial (kF^2k)/\partial B_y= 2 B_y k^{\prime 2}\sin \phi$. If
$lim B_y \to 0$, the second term vanishes. The claimed
``perpendicular component" vanishes in any case.

 Let us return to (\ref{magneticmom}).
 We may write it as a particular case of a more general problem, say, $\omega= f(B, g^{i}(B),
h^{j})$, where  $g^{i}(B)$ are arbitrary functions of $B$, and
$h^{j}$ are scalars independent of $B$ . It is easily shown that
\begin{equation}
\frac{\partial \omega}{\partial \textbf{B}}=\frac{\partial
f}{\partial B}\frac{\partial B}{\partial \textbf{B}}+ \frac{\partial
 f}{\partial g^{i}}\frac{\partial g^{i}}{\partial B}\frac{\partial B}{\partial
 \textbf{B}}=(\frac{\partial
f}{\partial B}+ \frac{\partial
 f}{\partial g^{i}}\frac{\partial g^{i}}{\partial
 B})\textbf{n}_{\parallel}, \label{mostgen}
\end{equation}
where the sum over $i$ is understood. We see that $\frac{\partial
\omega}{\partial \textbf{B}}$ is a vector parallel to $\textbf{B}$.
(Notice that the operator $\partial/\partial
\textbf{B}=\textbf{n}_{\parallel}\partial/\partial B$, and under a
rotation of coordinates $n_{\parallel k}^{\prime}=R_{kj}
n_{\parallel j}$). The only assumption done in (\ref{mostgen}) is
that $B=|\textbf{B}|$. We conclude that no perpendicular component
exists.

If, however, we consider some tensor  function, for instance, of the
dyadic $\textbf{B}\textbf{c}$, where $\textbf{c}$ is a vector non
parallel to $\textbf{B}$, a linear tensor function $\textbf{t}=a
\textbf{B}\textbf{c}$, it would lead to
\begin{equation}
\frac{\partial \textbf{t}}{\partial \textbf{B}}=a \textbf{c}.
\end{equation}
Thus, the vector $\frac{\partial \textbf{t}}{\partial \textbf{B}}$
is directed along $\textbf{c}$, not along $\textbf{B}$ and may have
a component perpendicular to $\textbf{B}$. But for the photon in
magnetized vacuum, we are dealing with a different problem, since
$\omega$ is not a tensor, and \textit{it is not} a function of any
dyadic, but of the scalars $k_{\parallel}^2$, $z_2$, $B$.

\subsection{Physical criticism}
 For the transparency region, ($\omega< 2m$) the photon magnetic moment is
due to the vacuum magnetization arising from the electron-positron
pairs. The dynamics of observable electrons and positrons was
discussed  in \cite{Johnson}, and these results are valid for
virtual pairs of vacuum. All symmetry and conservation properties
are valid for vacuum pairs, in agreement to the content of a basic
theorem due to Coleman \cite{Coleman} which states that \textit{the
invariance of the vacuum is the invariance of the world}.

For electrons and positrons physical quantities are invariant only
under rotations around $x_3$ or displacements along it
\cite{Johnson}. This means that conserved quantities, whose
operators commute with the Hamiltonian, are all parallel to
$\mathbf{B}$, as angular momentum and spin components
$J_3$,$L_3$,$s_3$ and the linear momentum $p_3$. By using units
$\hbar=c=1$, the energy eigenvalues for $e^{\pm}$ are
$E_{n,p_3}=\sqrt{p_3^2+m^2+ eB(2n+1+s_3)}$ where $s_3=\pm 1$ are the
spin eigenvalues along $x_3$ and $n=0,1,2..$ are the Landau quantum
numbers. In other words, the transverse squared Hamiltonian $H_t^2$
eigenvalues are $E_{n,p_3}^2-p_3^2=eB(2n+1+s_3)$ is quantized as
integer multiples of $eB$.

The transverse squared Hamiltonian $H^2_t=(2eB)(J_z + eB r_0^2/2)$,
has eigenvalues are $(2n+ 1+s_3)eB$. Here $r_0^2$ is the squared
center of the orbit coordinates operator, with eigenvalues
$(2l+1)/eB$, and the eigenvalues of $J_z$ are $n-l + s_3/2$. Thus,
the energy is degenerate with regard to the quantum number $l$, or
either, with regard to the momentum $p_y$ or the orbit's center
coordinate $x_0=p_y/eB$.

 The magnetic moment operator $M$ is the sum of two terms one of
which \cite{Johnson} is not a constant of motion but its
\textit{quantum average} vanishes. Its expectation value is $\bar
M=-<\partial H/\partial B>=-\partial E_{n,p_3}/\partial B$, where
$H$ is the Dirac Hamiltonian. Then $\bar M=-(E^2-p_3^2-m^2)/B E$,
and is the modulus of a vector parallel to $B$ for negative energy
states, antiparallel to B for positive energy states, and $\bar
\mathbf{M}=\bar M \mathbf{n}_{\parallel}$.  From the previous
paragraph, we see that \textit{there is no any linear relation
between $\bar M$ and $J_3$ as  in non-relativistic quantum
mechanics}, since $\bar M$ is a nonlinear function of the
eigenvalues of $J_z$ and $r_0^2$. \textit{There is no room for an
electron magnetic moment component orthogonal to $B$.}

On the opposite, a  photon magnetization is expected to be also
aligned along $\textbf{B}$. Let us return to some results of
\cite{EliPRD}.  It may be conceived for the photon a $B$- dependent
angular momentum produced by its transverse momentum interacting
with virtual charged fermions. We may define a quantity with
dimensions of length as the modulus of a vector $\textbf{r}_0$
located in the plane orthogonal to $\textbf{B}$. By multiplying it
by the momentum component $\textbf{k}_{\perp}$, we obtain a quantity
which we may call ``magnetized photon angular momentum"
$\textbf{J}_{\gamma}= \textbf{r}_0 \times \textbf{k}_{\perp}$, which
is parallel to $J_3$. It increases proportional to $\sqrt{eB}$, as
we shall see below.

Let us write from \cite{EliPRD} the basic equation
$\omega^{(i)2}=\vert\textbf{k}\vert^2+\mathfrak{M}^{2(i)}\left(z_2,B\right)$.
The second term at the right depends only on
$\textbf{k}_{\perp}^2\equiv z_2$, which implies a contribution to
angular momentum along $\textbf{B}$, and no contribution to the
direction perpendicular to it. The photon magnetic moment comes from
$\boldsymbol{\mu}_{\gamma}^{(i)}=-\partial \omega/\partial
\textbf{B}= -(\partial
\mathfrak{M}^{2(i)}\left(z_2,B\right)/\partial \textbf{B})/2\omega$.
We have, in the low frequency, low magnetic field limit, from
\cite{EliPRD}(we recall that $b=B/B_c$, where $B_c$ is the Schwinger
critical field)
\begin{equation}
\boldsymbol{\mu}_{\gamma}^{(2)}= \frac{14 \alpha z_2 }{45 \pi B_c
\omega} \left(b -\frac{52 b^3}{49}\right)\textbf{n}_{\parallel}
\end{equation} By returning to standard units $\hbar, c$, and defining the
vector $\textbf{r}_0$ orthogonal to $\textbf{B}$ of modulus
$r_0=\sqrt{\frac{\hbar c B}{e B^2_c}}$, we can write
\begin{equation}
\boldsymbol{\mu}_{\gamma}^{(2)}= \frac{14 \alpha}{45 \pi }\frac{e
\hbar c}{\omega}\frac{|\textbf{J}_{\gamma}|^2}{\hbar^2} \left(1
-\frac{104 b^2}{49}\right)\textbf{n}_{\parallel}.
\end{equation}
This (approximate) expression suggests a quadratic dependence of
$\boldsymbol{\mu}_{\gamma}^{(2)}$ with regard to the $B$-dependent
angular momentum $\textbf{J}_{\gamma}=\textbf{r}_0 \times
\textbf{k}_{\perp}$. A similar expression can be obtained for
$\boldsymbol{\mu}_{\gamma}^{(3)}$. Higher powers on $z_2$ in the
expansion of $\kappa^{(2)}$ would lead to higher powers of
${J}_{\gamma}^2/\hbar^2$.

 But
there is no any rotational symmetry to associate an hypothetical
angular momentum component orthogonal to $\textbf{B}$ (the component
$\textbf{r}_0 \times \textbf{k}_{\parallel}$ does not appear in our
formulae). We recall also that for the case of propagation
perpendicular to $\textbf{B}$, the modes $a^{(2,3)}$ are plane
polarized \cite{shabad2}. This means that they are a superposition
of waves of opposite helicity \cite{EliPRD}, which implies
superposition of spin states $S= \pm 1$. This is valid for
nonparallel propagation, which results from a Lorentz boost along
$\textbf{B}$ of the case of perpendicular propagation. In other
words, this means not well defined angular momentum states
orthogonal to $\textbf{B}$. All this precludes any photon magnetic
moment component orthogonal to $B$.

 \section{Acknowledgement} The authors would like to thank ICTP Office
 of External Activities for its support to the present research under
 NET-35.

\end{document}